\documentclass[10pt,A4]{iopart}
\usepackage{iopams}
\expandafter\let\csname equation*\endcsname\relax 
\expandafter\let\csname endequation*\endcsname\relax 
\usepackage{amsmath}
\usepackage{amssymb}

\usepackage{graphicx,color}           

\def\g{{\ {\rm g}}}                          

\def\P{{\ {\rm P}}}                         

\def\ms{{\ {\rm ms}}}                       

\def\Schrodinger{{Schr\"odinger\ }}
\def\pprime{{\prime\prime}}                     

\def\Er{E_R}                            
\def\El{E_L}                            
\def\kr{k_R}                            
\def\kl{k_L}                            
\def\Rb87{^{87}\text{Rb}}                     
\def\Na23{^{23}\text{Na}}                     
\def\K41{^{41}\text{K}}                     
\def\Li6{^{6}\text{Li}}                       
\def\Li7{^{7}\text{Li}}                       

\def\ex{{\mathbf e}_x}                            
\def\ey{{\mathbf e}_y}                            
\def\ez{{\mathbf e}_z}                            

\DeclareMathAlphabet\mathbfcal{OMS}{cmsy}{b}{n}
\def\Ageo{{\mathbfcal A}}                            
\def\Bgeo{{\mathbfcal B}}                            

\def\bra#1{\mathinner{\langle{#1}|}}
\def\ket#1{\mathinner{|{#1}\rangle}}

{\catcode`\|=\active
  \gdef\Braket#1{\left<\mathcode`\|"8000\let|\BraVert {#1}\right>}}
\def\BraVert{\egroup\,\mid@vertical\,\bgroup}

\begin{document}

\title{Flux lattices reformulated}

\author{G.~Juzeli\={u}nas$^1$ and I.~B.~Spielman$^2$}
\address{$^1$Institute of Theoretical Physics and Astronomy, Vilnius University, A. Go\v{s}tauto 12, Vilnius LT-01108, Lithuania}
\address{$^2$Joint Quantum Institute, National Institute of Standards and Technology, and University of Maryland, Gaithersburg, Maryland, 20899, USA}
\ead{Gediminas.Juzeliunas@tfai.vu.lt}
\ead{ian.spielman@nist.gov}

\date{\today}

\begin{abstract}
We theoretically explore the optical flux lattices produced for ultra-cold atoms
subject to laser fields where both the atom-light coupling and the effective detuning
are spatially periodic. We analyze the geometric vector potential and the magnetic
flux it generates, as well as the accompanying geometric scalar potential. We show how
to understand the gauge-dependent Aharonov-Bohm singularities in the vector potential, and  calculate the continuous magnetic flux through the elementary cell in terms of these singularities.  The analysis is illustrated with a square optical flux lattice. We conclude with an explicit laser configuration yielding such a lattice using a set of five properly  chosen beams with two counterpropagating pairs (one along the $x$ axes and the other $y$ axes), together with a single beam along the $z$ axis.  We show that this lattice is not phase-stable, and identify the one phase-difference that affects the magnetic flux.  Thus armed with realistic laser setup, we directly compute the Chern number of the lowest Bloch band to identify the region where the non-
zero magnetic flux produces a topologically non-trivial band structure.
\end{abstract}


\maketitle

\section{INTRODUCTION}

Atomic quantum gases are systems where condensed matter and
the atomic physics meet.  Cold atomic gases exhibit a number condensed
matter phenomena~\cite{Lewenstein2007,Ketterle2007,Bloch2008a,Giorgini2008},
such as the superfluid-Mott transition~\cite{Greiner2002}, Berezinskii-Kosterlitz-Thouless
superfluidity~\cite{Hadzibabic2006}, and the Bose-Einstein condensation to Bardeen-Cooper-Schrieffer (BCS) crossover~\cite{Greiner2003a,Zwierlein2004}.  Because the atoms comprising these quantum
gases are electrically neutral, no vector potentials
affect their center of mass motion.  Such a vector potentials might provide the Lorentz force essential for the magnetic phenomena
in solids, such as the quantum Hall effect~\cite{Klitzing1986}.
The standard way to produce an artificial magnetic field is to rotate
an atomic cloud leading to a non-trivial vector potential in the rotating
frame of reference~\cite{Cooper2008,Fetter2009}. The various proposed schemes to create effective
magnetic field for ultra-cold atoms without rotation~\cite{Dalibard2011} can be divided into two categories.

The first relies on a primary optical lattice which traps atoms at
its sites. The magnetic flux is created by inducing asymmetric tunneling between lattice sites, so that atoms acquire a non-zero phase after completing a closed loop along a plaquette~\cite{Jaksch2003,Mueller2004,Sorensen2005,Osterloh2005,Lim2008,Gerbier2010,Kolovsky2011,Demler2010,Bloch2011,Sengstock2012}.  Such asymmetries can be induced by laser-assisted tunneling~\cite{Jaksch2003,Mueller2004,Sorensen2005,Osterloh2005,Gerbier2010,Bloch2011,Ruostekoski:2002}
or using time-dependent lattices~\cite{Sorensen2005,Lim2008,Kolovsky2011,Demler2010,Sengstock2012}. 

The second group of proposals is based on the concept of geometric
gauge potentials which are encountered in many areas of physics~\cite{Berry1984,Jackiw1988,Moody1986,Bohm92JMP,Zee1988,Shapere1989,Mead1992,Bohm2003,Xiao2010}.
In atomic gases, the geometric vector and scalar
potentials were first considered in the late 90's for atoms interacting
with the laser fields~\cite{Dum1996,Visser1998,Dutta1999}, where the atoms are ``dressed'' by
laser beams. The resulting position dependence of the dressed internal
states leads to geometric vector and scalar
potentials. The method can provide a non-zero effective magnetic field
using non-trivial spatial arrangements of laser fields~\cite{Juzeliunas2004,Juzeliunas2005b,Ruseckas2005,Zhang-Sun2005,Juzeliunas2006,Zhu2006,Gunter2009,Cooper2010}
or position-dependent detuning of the atom-light coupling~\cite{Spielman2009,Lin2009b,Lin2009a}.
In these approaches the magnetic flux through the atomic cloud scales
linearly with the cloud's extent~\cite{Juzeliunas2006,Gunter2009,Spielman2009,Lin2009b}, not it's area.
For large systems this is a major obstacle in reaching the sizable magnetic fluxes required to achieve the fractional Hall effect~\cite{Laughlin1999}. 

A new class of geometric potentials termed ``flux lattices'' were recently shown to yield a magnetic flux  proportional to the surface area of the atomic cloud~\cite{Cooper2011a},
see also~\cite{Spielman2011}.  In this proposal, a
two-level atom was coupled to a spatially periodic laser field where both the atom-light coupling and the detuning term were oscillatory.  This approach simultaneously generates a non-staggered magnetic
flux along with a lattice potential thus providing an optical
flux lattice~\cite{Cooper2011a}. 

The vector potential $\mathbf{A}$ plays an important role in the
quantum physics~\cite{Aharonov1959}. It is featured in the Pearls
substitution~\cite{Peierls1933,Luttinger1951,Hofstadter1976} widely used in the
tight binding models in solids to describe the motion of charged particles in the magnetic field.
Specifically, the tunneling matrix element between the lattice sites
$\mathbf{r}_{A}$ and $\mathbf{r}_{B}$ acquires the Pearls phase
factor proportional to $\int_{\mathbf{r}_{A}}^{\mathbf{r}_{B}}\mathbf{A}\cdot d\mathbf{r}$.
Similar Pearls phase factors emerge also in the tunneling matrix elements between 
the sites of an optical lattice for electrically neutral ultracold atoms affected by the artificial magnetic field. 

It is instructive that the geometric vector potential for the optical flux lattices contains gauge-dependent
singularities.  To avoid these singularities, Ref.~\cite{Cooper2011a}
concentrated on the magnetic flux rather on the vector potential.  Here we explore optical flux lattices explicitly in terms of their geometric vector potential (which generates the magnetic flux) and an accompanying geometric scalar
potential.  We show how to understand the gauge-dependent Aharonov-Bohm (AB) singularities~\cite{Aharonov1959} appearing
in this vector potential and we calculate the continuous magnetic flux through the elementary
cell in terms the singularities.

Subsequently, we analyze a square optical flux lattice and describe
a way of creating it using a combination of state-dependent lattice potentials and Raman transitions.  A related setup
proposed recently by Cooper and Dalibard aimed at producing
triangular and hexagonal optical flux lattices used three coplanar lasers intersecting
at 120 degrees, with an additional beam normal to the plane spanned by the first three~\cite{Cooper2011}.  Here we present an explicit laser configuration yielding a square flux lattice 
and directly compute the Chern number of the lowest Bloch band. We identify the region where the non-zero magnetic flux produces a topologically non-trivial band structure for this lattice.

\section{Hamiltonian and its eigenstates\label{sec:Hamiltonian-and-its}}

Before focusing on a specific physical system, we begin by considering the very general problem of a multi-level atom moving in the presence of a spatially inhomogenous coupling Hamiltonian (for example produced by a combination of optical and magnetic fields).  The Hamiltonian describing such a combined internal and center of mass motion is
\begin{equation}
\hat{H}=\left[\frac{\mathbf{p}^{2}}{2m}\,+U(\mathbf{r})\right]\hat{1}\,+\hat{M}(\mathbf{r}),\label{eq:hamiltonian1}
\end{equation}
where $\mathbf{p}=-i\hbar\boldsymbol{\nabla}$ is the atomic momentum;
$\hat{1}$ is the identity operator; $U(\mathbf{r})$ is a state-independent
``scalar'' potential; and $\hat{M}$ is the state-dependent part of
the Hamiltonian.  Here we focus on the case where the atom affected by the light fields behaves
like a spin in a magnetic field, so the state dependent Hamiltonian
$\hat{M}(\mathbf{r})$ is 
\begin{equation}
\hat{M}(\mathbf{r})=\boldsymbol{\Omega}\cdot\hat{\boldsymbol{F}}\equiv\Omega\hat{F}_{\Omega},\label{eq:U}
\end{equation}
where the vector $\boldsymbol{\Omega}\equiv\boldsymbol{\Omega}(\mathbf{r})=(\Omega_{x}(\mathbf{r}),\Omega_{y}(\mathbf{r}),\Omega_{z}(\mathbf{r}))$ 
describes the spatially dependent coupling between the atomic internal states,
$\Omega(\mathbf{r})=\left|\boldsymbol{\Omega}(\mathbf{r})\right|$
being the total coupling strength; $\hat{\boldsymbol{F}}=\left({\hat F}_{x},{\hat F}_{y},{\hat F}_{z}\right)$
is a vector operator satisfying the angular momentum algebra (to be
referred to as the spin operator); and $\hat{F}_{\Omega}\equiv\hat{F}_{\Omega(\mathbf{r})}$ is the spatially dependent projection of $\hat{\boldsymbol{F}}$ along  $\boldsymbol{\Omega}$.  The position dependence of the Hamiltonian $\hat{M}(\mathbf{r})$ therefore originates
from the position-dependent atom-light interaction through the coupling vector $\boldsymbol{\Omega}(\mathbf{r})$: a rapidly varying effective magnetic field.    The physical implementation of such a Hamiltonian -- physically equivalent to the Zeeman effect for a spatially dependent magnetic field -- will be discussed in Sec.~\ref{sec:alkali}.

\begin{figure}
\begin{centering}
\includegraphics[width=3.3in]{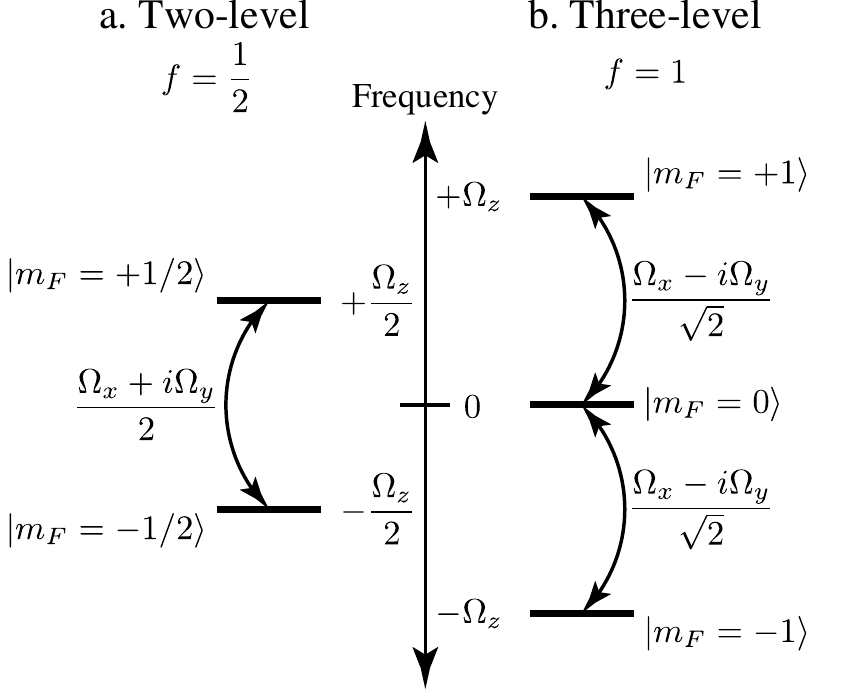}
\caption[Two and three level schemes]{\label{fig:fig1} Traditional representation of $\hat M = \boldsymbol{\Omega}\cdot\hat{\boldsymbol{F}}$ coupling scheme for (a) total angular momentum $f=1/2$ and (b) total angular momentum $f=1$.  The bare atomic states are labeled by $\ket{m_F}$, and in both cases are detuned from each other by a frequency $\Omega_{z}$.  These levels are coupled with strength proportional to $\Omega_{x}\pm i\Omega_{y}$.}
\end{centering}
\end{figure}

Figure~\ref{fig:fig1}a depicts the (quasi-)spin-$1/2$ case, where $\hbar\Omega_{z}$ is the light-induced detuning between the two internal atomic states $\ket{m_F = \pm1/2}$, and $\hbar\left(\Omega_{x}\pm i\Omega_{y}\right)/2$ is the transition
matrix element coupling the two states together. In what follows, we do not restrict
ourselves to the spin-$1/2$ case: for $N$ internal states $\left\{\ket{m_F}|m_F=-f,\,f+1,\,\ldots\,f\right\}$, the quantity $f=(N-1)/2$ is the total angular momentum quantum number.  The spin-$1$ case depicted in Fig.~\ref{fig:fig1}b, has $f=1$ and $N=3$.  We emphasize that the atomic states $\ket{m_F}$ do not necessarily represent the true spin states. They can be the atomic internal states of arbitrary origin, provided that the operator $\hat{F}$ featured in the atomic Hamiltonian obeys the angular momentum algebra.  

\subsection{Diagonalization via a unitary transformation}

\begin{figure}
\begin{centering}
\includegraphics[width=3.3in]{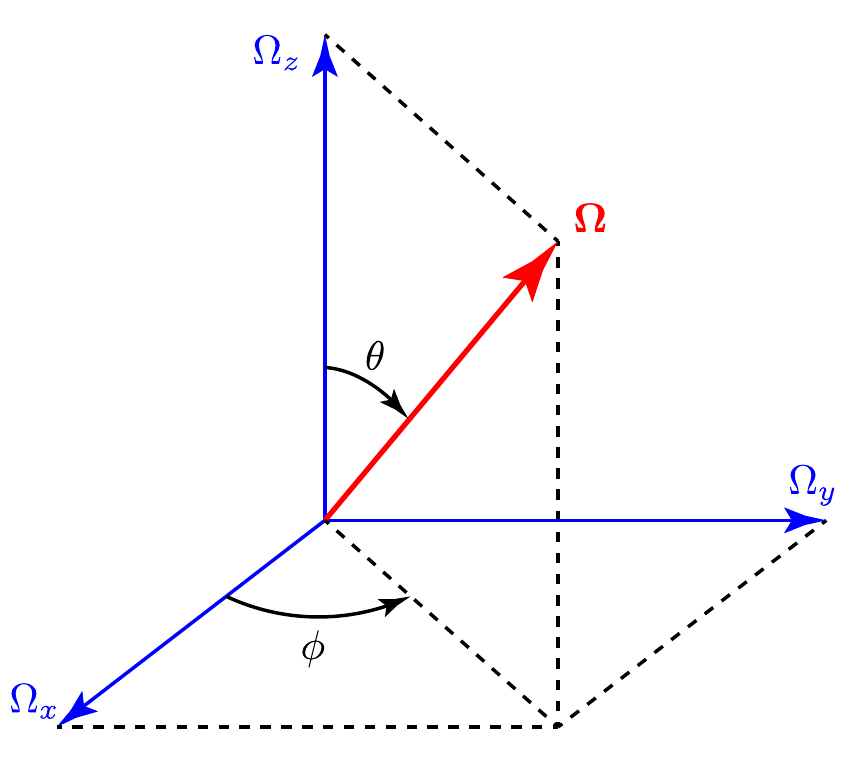} 
\caption{Representation of the coupling vector $\boldsymbol{\Omega}=(\Omega_{x},\Omega_{y},\Omega_{z})$
in terms of the spherical angles $\theta$ and $\phi$ .}
\label{fig:fig2-Omega} 
\end{centering}
\end{figure}

The projected momentum operator $\hat{F}_{\Omega}$ entering the coupling Hamiltonian
$\hat{M}$ can be related to the eigenstates of ${\hat F}_{z}$ via a unitary transformation ${\hat S}_{\Omega}$ where
\begin{equation}
\hat{F}_{\Omega}={\hat S}_{\Omega}\hat{F}_{z}{\hat S}_{\Omega}^{-1}\,,\label{eq:F-Omega}
\end{equation}
with
\begin{equation}
{\hat S}_{\Omega}=e^{-i\hat{F}_{z}\phi/\hbar}e^{-i\hat{F}_{y}\theta/\hbar}e^{i\hat{F}_{z}\phi/\hbar}.\label{eq:S-Omega}
\end{equation}
We  parametrize the coupling vector $\boldsymbol{\Omega}=(\Omega_{x},\Omega_{y},\Omega_{z})$
in terms of the spherical angles 
\begin{align}
\tan\phi&=\frac{\Omega_{y}}{\Omega_{x}}, &&{\rm and} & \qquad\cos\theta&=\frac{\Omega_{z}}{\Omega}\label{eq:tan-phi--theta}
\end{align}
shown in Fig.~\ref{fig:fig2-Omega}.

The operators ${\hat {\mathbf F}}^2$ and ${\hat F}_{z}$ have eigenstates $\left|m_{F}\right\rangle \equiv\left|f,m_{F}\right\rangle $ identified by the total angular momentum $f$ and its $\ez$ projection $m_{F}$ with eigenvalues
\begin{align*} 
{\hat {\mathbf F}}^2\ket{m_{F}} &=\hbar^2 f(f+1)\ket{m_{F}}\\
{\hat F}_{z}\ket{m_{F}} &=\hbar m_{F}\ket{m_{F}}.
\end{align*}
Multiplying the last equation by ${\hat S}_{\Omega}$, one has
\begin{align}
\hat{F}_{\Omega}\left|\tilde{m}_{F}\right\rangle &=\hbar m_{F}\left|\tilde{m}_{F}\right\rangle, && {\rm where} &
\left|\tilde{m}_{F}\right\rangle ={\hat S}_{\Omega}\left|m_{F}\right\rangle
.\label{eq:eigenstates-m_F}
\end{align}
Therefore the coupling Hamiltonian $\hat{M}=\Omega\hat{F}_{\Omega}$
has a set of position-dependent eigenstates $\left|\tilde{m}_{F}\right\rangle \equiv\left|\tilde{m}_{F}(\mathbf{r})\right\rangle $
related to the eigenstates of $\hat{F}_{z}$ via the position-dependent
unitary transformation ${\hat S}_{\Omega}\equiv {\hat S}_{\Omega}(\mathbf{r})$.
Using Eq.~\eqref{eq:S-Omega} for ${\hat S}_{\Omega}$ and the fact that $\left|m_{F}\right\rangle $
is an eigenstate of $\hat{F}_{z}$, one arrives at the eigenstates
\begin{align}
\left|\tilde{m}_{F}\right\rangle &=e^{i\left(m_{F}-\hat{F}_{z}/\hbar\right)\phi}e^{-i\hat{F}_{y}\theta/\hbar}\left|m_{F}\right\rangle,
\end{align}
and energies
\begin{align}
V_{m_F}&=\hbar m_{F}\Omega\label{eq:eigenstates-F_z-tilde-1}.
\end{align}
of the coupling Hamiltonian $\hat{M}$.  Here, $V(\mathbf{r})\equiv V_{m_F}$ is the position dependent energy of the local eigenstate $\left|\tilde{m}_{F}(\mathbf{r})\right\rangle$.  Interestingly, similar kinds of eigenstates give rise to artificial gauge potential terms describing the rotation of diatomic molecules~\cite{Moody1986,Bohm92JMP} and the physics of atomic collisions~\cite{Zygelman:1987,Zygelman:1990}.

\section{Gauge potentials}

For an atom subject to the Hamiltonian of Eq.~\eqref{eq:hamiltonian1}, the state-vector describing both it's internal and motional degrees of freedom can be expressed in the basis of dressed states 
\[
\left|\Psi(\mathbf{r},t)\right\rangle =\sum_{m_{F}^{\prime}}\psi_{m_{F}^{\prime}}(\mathbf{r},t)\ket{\tilde{m}_{F}^{\prime}},
\]
where $\psi_{m_{F}^{\prime}}(\mathbf{r},t)$ is a wave-function
describing the atom's motion in the basis of local eigenstates $\left|\tilde{m}_{F}^{\prime}\right\rangle \equiv\left|\tilde{m}_{F}^{\prime}(\mathbf{r})\right\rangle $.  We are interested
in the situation where $\Omega>0$, so the local
eigenstates $\ket{\tilde{m}_{F}^{\prime}}$ are everywhere non-degenerate. 

If an atom is prepared in one of these dressed states with $m_{F}^{\prime}= m_{F}$, and it's characteristic kinetic energy is small compared to the $\Delta E=\hbar\Omega$ the energy difference between
adjacent spin states, the internal state of the atom will adiabatically follow the
dressed state   $\left|\tilde{m}_{F}\right\rangle $ as the atom moves, and contributions due to other states
with $\tilde{m}_{F}^{\prime}\ne m_{F}$ can be neglected.  Projecting
the full \Schrodinger equation $i\hbar\ket{\dot{\Psi}(\mathbf{r},t)} =\hat{H}\ket{\Psi(\mathbf{r},t)}$
onto the selected internal eigenstate $\left|\tilde{m}_{F}\right\rangle $ yields a reduced \Schrodinger
equation for the atomic center of mass motion  
$i\hbar\dot{\psi}_{m_{F}}(\mathbf{r},t) =H\psi_{m_{F}}(\mathbf{r},t)$ with an effective Hamiltonian
\begin{equation}
H=\frac{\left[\mathbf{p}-\Ageo(\mathbf{r})\right]^{2}}{2m}\,+U(\mathbf{r})+V(\mathbf{r})+W(\mathbf{r}).\label{eq:Eq-mot-general}
\end{equation}
In this expression, the geometric vector
\begin{align}
\Ageo\equiv\Ageo(\mathbf{r})&=i\hbar\bra{\tilde{m}_{F}}\boldsymbol{\nabla}\ket{\tilde{m}_{F}}\label{eq:A-def}
\end{align}
and scalar 
\begin{align}
W\equiv W(\mathbf{r})&=\frac{\hbar^{2}}{2m}\sum_{{\tilde m}_{F}^\pprime\ne {\tilde m}_{F}}\left|\bra{\tilde{m}_{F}^{\pprime}}\boldsymbol{\nabla}\ket{\tilde{m}_{F}} \right|^{2}\label{eq:W-def}
\end{align}
potentials appear due to the position dependence of the atomic dressed states.  
The vector potential can be interpreted as the average center of mass momentum of the selected internal state $\left|m_{F}\right\rangle \equiv\left|m_{F}\left(\mathbf{r}\right)\right\rangle $.  The scalar potential $W(\mathbf{r})$ emerges due to the elimination of the remaining atomic internal states.
It represents the kinetic energy of the oscillatory micromotion~\cite{Aharonov1992,Cheneau2008} due
to the tiny transitions to the eliminated states $\left|\tilde m_F^{\pprime}\right\rangle \equiv\left|\tilde m_F^{\pprime}\left(\mathbf{r}\right)\right\rangle$ with ${\tilde m}_{F}^\pprime\ne {\tilde m}_{F}$. 

Equation~\eqref{eq:Eq-mot-general} contains three distinct scalar potentials: (a) the state independent potential $U(\mathbf{r})$ featured in the initial Hamiltonian \eqref{eq:hamiltonian1}, which we shall call $U(\mathbf{r})$ the ``scalar light shift;'' (b) the ``adiabatic scalar potential'' $V(\mathbf{r})\equiv V_{m_F}$ arising from spatial variations in the magnitude $\Omega(\mathbf{r})$; and (c) the ``geometric scalar potential'' $W(\mathbf{r})$ described above.  All three contribute to the potential energy of atoms in the dressed state basis. 

\subsection{Vector and scalar potentials}

Using Eqs.~\eqref{eq:eigenstates-F_z-tilde-1}, the matrix elements featured in the vector and scalar potentials are
\begin{align}
\left\langle \tilde{m}_{F}^{\pprime}\right|\boldsymbol{\nabla}\left|\tilde{m}_{F}\right\rangle &= \frac{i}{\hbar}e^{i\left(\tilde{m}_{F}-\tilde{m}_{F}^{\pprime}\right)\phi}\bra{m_{F}^{\pprime}}\left[\left(\hbar m_{F}-\hat{\tilde{F}}_{z}\right) \nabla\phi-\hat{F}_{y} \nabla\theta\right]\ket{m_{F}},\label{eq:nabla-action-matrix-element}
\end{align}
with $\hat{\tilde{F}}_{z}=\exp(i\hat{F}_{y}\theta/\hbar)\hat{F}_{z}\exp(-i\hat{F}_{y}\theta/\hbar)=\hat{F}_{z}\cos\theta+\hat{F}_{x}\sin\theta$.  Using the identities $\left\langle m_{F}\right|\hat{F}_{x}\left|m_{F}\right\rangle =\left\langle m_{F}\right|\hat{F}_{y}\left|m_{F}\right\rangle =0$ and $\left\langle m_{F}\right|\hat{F}_{z}\left|m_{F}\right\rangle =m_{F}$, Eqs.~\eqref{eq:A-def}-\eqref{eq:nabla-action-matrix-element} provide the vector potential
\begin{equation}
\Ageo(\mathbf{r})=\hbar m_{F}\left(\cos\theta-1\right)\boldsymbol{\nabla}\phi.\label{eq:vect}
\end{equation}
The vector potential~\eqref{eq:vect} is maximum in magnitude when $m_{F}=\pm f$, and is zero for $m_{F}=0$. For $m_{F}=\text{1/2}$, Eq.~\eqref{eq:vect} reduces to the result presented in Ref.~\cite{Dalibard2011}. 

To determine the scalar potential, we need the off diagonal
matrix elements of $\left\langle \tilde{m}_{F}^{\pprime}\right|\boldsymbol{\nabla}\left|\tilde{m}_{F}\right\rangle $
which are
\begin{align}
\left\langle \tilde{m}_{F}^{\pprime}\right|\boldsymbol{\nabla}\left|\tilde{m}_{F}\right\rangle &= -\frac{i}{\hbar}e^{i\left(\tilde{m}_{F}-\tilde{m}_{F}^{\pprime}\right)\phi}\bra{ m_{F}^{\pprime} }\left(\hat{F}_{x} \sin\theta\nabla\phi+\hat{F}_{y} \nabla\theta\right)\ket{m_{F}},\label{eq:nabla-action-matrix-element-1}
\end{align}
for $\tilde m_{F}^{\pprime}\ne \tilde m_{F}$.  Combining Eqs.~\eqref{eq:W-def} and \eqref{eq:nabla-action-matrix-element-1}, and using the completeness relation, we arrive at the geometric scalar potential 
\begin{align}
W(\mathbf{r})&=\frac{\hbar^{2}}{4m}g_{f,m_F}\left[\sin^2\theta\left(\nabla\phi\right)^{2}+\left(\nabla\theta\right)^{2}\right],\label{eq:W-result}
\end{align}
with
\begin{align}
g_{f,m_F}=f(f+1)-m_{F}^{2},\label{eq:g}
\end{align}
where we made use of: $\bra{m_{F}}\hat{F}_{x}^{2}\ket{m_{F}} = \bra{m_{F}}\hat{F}_{y}^{2}\ket{m_{F}} =g_{f,m_F}/2$.
In particular, for $f=1/2,$ one has $g_{f,m_F}=1/2$, and
the scalar potential reduces to that presented in Ref.~\cite{Dalibard2011}.
Generally $W(\mathbf{r})$ depends both on the total
spin $f$ and also on its projection $m_{F}$. For instance, for $f=1,$
one has $g_{f,m_F}=2-m_{F}^{2}$, showing that $W({\mathbf r})$ is maximum for $m_{F}=0$
and is a half of that for $m_{F}=\pm1$.

\subsection{Alternative gauge}

Because each of the local eigenstates can be assigned an arbitrary position dependent phase $\ket{{\tilde m}_F^\prime}=\exp\left[i\varphi_{{\tilde m}_F}\left({\mathbf r}\right)\right]\ket{{\tilde m}_F}$, effecting a state-dependent gauge transformation, the vector potential $\Ageo(\mathbf{r})$ in Eq.~\eqref{eq:vect} is not unique.  For example, when 
\begin{align}
\ket{{\tilde m}_F^{\prime}}&=e^{-2i\phi_{{\tilde m}_F}}\ket{{\tilde m}_F}
\end{align}
the initial vector potential given by Eq.~(\ref{eq:vect}) becomes
\begin{align*}
\Ageo^{\prime}(\mathbf{r})&=\hbar m_{F}\left(\cos\theta+1\right)\boldsymbol{\nabla}\phi.
\end{align*}
This seemingly esoteric change can have significant impact because the ${\nabla}\phi$ contribution may be singular when $\cos\theta=\pm1$ (near the $z$ axes, see Fig.~\ref{fig:fig2-Omega}) if the factors $\left(\cos\theta-1\right)$ or  $\left(\cos\theta+1\right)$ do not compensate the singularity by simultaneously going to zero.  The vector potential $\Ageo(\mathbf{r})$, may have singularities when $\theta=\pi$, but in the alternative gauge the vector potential $\Ageo^{\prime}(\mathbf{r})$ has potential singularities when $\theta=0$: {\it at spatially different points} than the initial gauge!  

\subsection{Magnetic flux}

The singularities in the vector potential correspond to AB type flux tubes (piercing the $\ex$-$\ey$ plane) each with an integer flux quantum.  Since the AB type flux containing an integer number flux quanta can not be observed~\cite{Aharonov1959}, the two vector potentials $\Ageo(\mathbf{r})$
and $\Ageo^{\prime}(\mathbf{r})$ are equivalent and produce the same effective magnetic field
\begin{equation}
\Bgeo(\mathbf{r})=\boldsymbol{\nabla}\times\Ageo(\mathbf{r})=\hbar m_F\boldsymbol{\nabla}\left(\cos\theta\right)\times\boldsymbol{\nabla}\phi.\label{eq:B}
\end{equation}
The gauge-dependent AB singularities (if any) present in the
vector potential must be absent in Eq.~\eqref{eq:B} for $\Bgeo(\mathbf{r})$.

It is convenient to represent the magnetic flux density in terms of
the unit vector $\mathbf{N}=\boldsymbol{\Omega}/\Omega$
\begin{equation}
\Bgeo(\mathbf{r})=-\hbar m_F\frac{\nabla N_{x}\times\nabla N_{y}}{N_{z}}\,,\label{eq:B-alter}
\end{equation}
Thus if $\Omega_{z}$ alternates the sign at the points where $\Omega_{x}=\Omega_{y}=0$, 
this might compensate the alternation of the sign
of $\nabla N_{x}\times\nabla N_{y}$ at these points, giving a non-zero
magnetic flux, such as the one given by Eq.~\eqref{eq:B-Square} below.
This shows the necessity to have an oscillating detuning $\Omega_{z}$
in addition to the oscillating coupling $\Omega_{x}+i\Omega_{y}$.

As we show in Sec.~\ref{sec:Scalar}, the geometric scalar potential $W({\mathbf r})$ contributes most significantly to the overall scalar potential $U({\mathbf r}) + V({\mathbf r}) + W({\mathbf r})$ at the maxima of the effective magnetic field where $\Omega_{x}+i\Omega_{y}=0$, and is zero at the points of the minimum magnetic flux where $\Omega_{z}=0$.

\subsection{Periodic atom-light coupling}

Given this general background, we now consider the case where the coupling vector $\boldsymbol{\Omega}=(\Omega_{x},\Omega_{y},\Omega_{z})$ is spatially periodic in the $\ex$-$\ey$ plane 
\begin{equation}
\boldsymbol{\Omega}(\mathbf{r}+\mathbf{r}_{n,m})=\boldsymbol{\Omega}(\mathbf{r})\,,\qquad\mathbf{r}_{n,m}=n\mathbf{a}_{1}+m\mathbf{a}_{2}\,,\label{eq:Omega-periodic}
\end{equation}
where $\mathbf{a}_{1}$ and $\mathbf{a}_{2}$ are the primitive vectors
defining a 2D lattice in the $\ex$-$\ey$ plane, with $\left\{n,m\right\}\in \mathbb{Z}$. In this case, both the atomic internal dressed states
$\ket{{\tilde m}_F({\mathbf r})}$ and the corresponding geometric potential $\Ageo(\mathbf{r})$ have the
same periodicity [usually the geometric scalar potential $W(\mathbf{r})$ has a twice smaller periodicity than the initial Hamiltonian and the geometric vector potential].   Due to the periodicity of the vector potential the
total flux over the elementary cell is zero
\begin{align}
\alpha &=\frac{1}{\hbar}\oint_{\rm cell}\!\Ageo\cdot d\mathbf{r}=\frac{1}{\hbar}\iint_{\rm cell}\!\Bgeo_{\rm tot}\cdot d\mathbf{S}=0,\label{eq:flux-alpha}
\end{align}
where $\Bgeo_{\rm tot}=\Bgeo(\mathbf{r})+\Bgeo_{\rm AB}(\mathbf{r})$ is the total magnetic flux density with contributions both from the continuous
(background) magnetic flux density $\Bgeo(\mathbf{r})$ and 
possibly a set of gauge-dependent singular fluxes of the AB type represented
by $\Bgeo_{\rm AB}(\mathbf{r})$. 

Thus it is strictly speaking impossible to produce a non-zero effective magnetic
flux $\alpha$ over the elementary cell using the periodic atom-light
coupling. However, this does not preclude  a non-staggered continuous
magnetic flux density $\Bgeo(\mathbf{r})$ over the elementary
cell as long as the vector potential contains (gauge-dependent) singularities
of the AB type carrying together a non-zero number of the Dirac flux
quanta. The AB singularities are associated with the points where
the $\Omega_{x}+i\Omega_{y}$ goes to zero and hence
$\cos\theta\rightarrow\pm1$.  Deducting these non-measurable gauge dependent singularities, the
remaining flux over the elementary cell can be non-zero
\begin{equation}
\alpha^{\prime}=\frac{1}{\hbar}\iint_{\rm cell}\!\Bgeo\cdot d\mathbf{S}=-\frac{1}{\hbar}\iint_{\rm cell}\!\Bgeo_{\rm AB}(\mathbf{r})\cdot d\mathbf{S}\,.\label{eq:flux-alpha-prime}
\end{equation}
The physical flux can hence be expressed in terms of the vector potential 
\begin{equation}
\alpha^{\prime}=-\frac{1}{\hbar}\sum\oint_{\rm singul}\!\Ageo\cdot d\mathbf{r}=-\frac{1}{\hbar}\sum\oint_{\rm singul}\Ageo^{\prime}\cdot d\mathbf{r}\,,\label{eq:flux-alpha-prime-1}
\end{equation}
where the summation is over the singular points of the vector potential
(emerging at $\cos\theta\rightarrow -1$ for $\Ageo$ and at $\cos\theta\rightarrow 1$ for ${\Ageo}^{\prime}$) around which the contour integration is carried out. In the neighborhood of each singular point (different for $\Ageo$ and ${\Ageo}^{\prime}$)  the vector potentials have a Dirac-string (the AB singularity) piercing the $\ex$-$\ey$ plane, giving
\begin{align}
\Ageo&\rightarrow -2\hbar m_F\nabla\phi, & {\rm and} && \Ageo^{\prime} &\rightarrow 2\hbar m_F\nabla\phi.\label{eq:A-substitution}
\end{align}
Thus each integral in Eq.~\eqref{eq:flux-alpha-prime-1} provides an integer number of the Dirac
flux quanta. To obtain a non-zero
flux $\alpha^{\prime}$, the sum of the singular contributions must
be non-zero. The flux is maximum if all these singular contributions
have the same sign, as is the case for the square flux lattice considered below.

To summarize, the optical flux
lattice contains a background non-staggered magnetic field $\Bgeo$
plus an array of gauge-dependent Dirac-string fluxes of the opposite
sign as compared to the background. The two types of fluxes compensate each other so the total magnetic flux
over an elementary cell is zero as is required from the periodicity
of the Hamiltonian. However, the Dirac-string fluxes are non-measurable
and hence must be excluded from any physical consideration. As a result, a non-staggered
magnetic flux over the optical flux lattice is possible.

\section{SQUARE OPTICAL FLUX LATTICE}

We now construct a simple model flux lattice generated by a spatially periodic coupling vector $\boldsymbol{\Omega}=(\Omega_x,\Omega_y,\Omega_z)$ with components
\begin{align}
\Omega_{x}&=\Omega_{\bot}\cos(x\pi/a),\nonumber \\
\Omega_{y}&=\Omega_{\bot}\cos(y\pi/a),\label{eq:Omega-xyz-Square} \\
\Omega_{z}&=\Omega_{\parallel}\sin(x\pi/a)\sin(y\pi/a).\nonumber
\end{align}
This coupling has period $2a$ along $\ex$
and $\ey$.  For simplicity, we define dimensionless coordinates $x^\prime = \pi x/a$, $y^\prime = \pi y/a$, and $z^\prime = \pi z/a$.  In the symmetric case, $\Omega_{\bot}=\Omega_{\parallel}$,
the scheme reduces to the one considered previously in Ref.~\cite{Cooper2011a}.
As will be discussed in Sec.~\ref{sec:Generation-fo-the}, 
the coupling vector in Eq.~\eqref{eq:Omega-xyz-Square} can be produced for alkali atoms
using five laser beams intersecting at right angles: a pair counterpropagating along $\ex$, a second pair counterpropagating along $\ey$, and a single beam propagating along $\ez$.

The total Rabi frequency resulting from Eq.~\eqref{eq:Omega-xyz-Square} is
\begin{equation}
\Omega=\sqrt{\Omega_{\parallel}^{2}+\left(\Omega_{\bot}^{2}-\Omega_{\parallel}^{2}\right)\left(f_{x}^{2}+f_{y}^{2}\right)+\Omega_{\parallel}^{2}f_{x}^{2}f_{y}^{2}},\label{eq:Omega-Square}
\end{equation}
where $f_{u}=\cos(u^\prime)$.  The resulting adiabatic energies $V_{\rm m_F}=\hbar m_F\Omega$ have periodicity
$a$, half that of the atom-light coupling. 

\begin{figure}
\begin{center}
\begin{tabular}{c}
\includegraphics[width=3.3in]{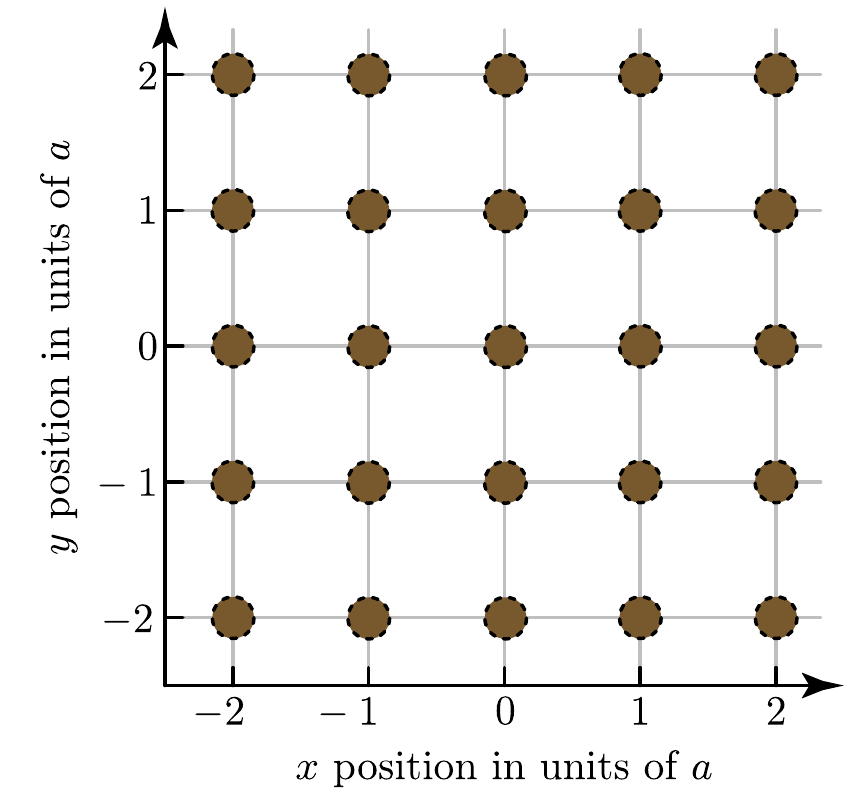}
\end{tabular}
\end{center}
\caption[Omega]{\label{fig:fig2}
Sites of the square optical flux lattice corresponding to the minima
of the adiabatic energy $V(x,y)$ for $m_F < 0$.}
\end{figure}

When $\Omega_{\bot}^2>\Omega_{\parallel}^2/2$, the minima of
the $m_F<0$ adiabatic scalar potential $V(x,y)$ are positioned at $x^\prime_{n}=\pi n$ and
$y^\prime_{m}=\pi m$ (brown dots in Fig.~\ref{fig:fig2}) where
$\Omega_{z}=0$. The energy maxima are positioned
at $x^\prime_{n,max}=\pi(n+1/2)$ and $y^\prime_{m,max}=\pi(m+/2)$ where the atom-light
coupling vanishes: $\Omega_{x}+i\Omega_{y}\rightarrow0$. Thus one
has
\begin{align}
E_{min}&=V(\pi n,\pi m)=\hbar m_F\sqrt{2}\Omega_{\bot},\nonumber\\
E_{max}&=V(\pi(n+1/2),\pi(m+/2) )=\hbar m_F\Omega_{\parallel}.\label{eq:E-Cooper-min-max}
\end{align}
In the vicinity of the energy maxima one has
\begin{align}
\Omega_{x}&\approx-\Omega_{\bot}\left(x^\prime-x^\prime_{n,max}\right)(-1)^{n},\nonumber\\
\Omega_{y}&\approx-\Omega_{\bot}\left(y^\prime-y^\prime_{m,max}\right)(-1)^{m},\label{eq:Omega-max} \\
\Omega_{z}&\approx\Omega_{\parallel}(-1)^{n+m}.\nonumber
\end{align}
Thus for odd (even) values of $n+m$ the angle $\phi$ rotates clockwise (anti-clockwise) around the singularities of the vector potential positioned at $x^\prime=x^\prime_{n,max}$ and $y^\prime=y^\prime_{n,max}$, whereas $\Omega_{z}$ alternates its sign when going from even to odd values of $n+m$.  This ensures a non-zero magnetic flux over the elementary cell when integrating the vector potential around its singular points in Eqs.~\eqref{eq:flux-alpha-prime-1} and \eqref{eq:A-substitution}.

\subsection{Magnetic flux}

Consider now the flux passing through the elementary cell with $x^\prime\in[0,2\pi)$ and $y^\prime\in[0,2\pi)$. 
The vector potential $\Ageo$ has Dirac-string singularities
for $\Omega_{z}=-\Omega_{\parallel}$ corresponding to odd values of $n+m$ in
Eq.~\eqref{eq:Omega-max}. Within the elementary cell these two
points are positioned at $\left(n=1,\,\, m=0\right)$ and $\left(n=0,\,\, m=1\right)$,
each providing $2m_F$ magnetic flux quanta.  In fact, integrating the vector potential around each singular
point, Eqs.~\eqref{eq:flux-alpha-prime-1}-\eqref{eq:A-substitution}
yield the background magnetic flux
\begin{equation}
\alpha^{\prime}=-\frac{1}{\hbar}\sum\oint_{\rm singul}\!\Ageo\cdot d\mathbf{r}=-8\pi m_F\,.\label{eq:flux-alpha-prime-square}
\end{equation}
In particular, for the spin-$1/2$ case ($m_F=1/2$)  a measurable continuous flux over the elementary cell accommodates
two Dirac quanta~\cite{Cooper2011a}. The
same gauge independent magnetic flux $\alpha^{\prime}$ is obtained using the alternate vector
potential $\Ageo^{\prime}$ which contains gauge dependent AB singularities at different points: $n=m=0$
and $n=m=1$, again each carrying $2m_F$ Dirac flux quanta. 

\begin{figure}[t!]
\begin{center}
\includegraphics[width=5.25in]{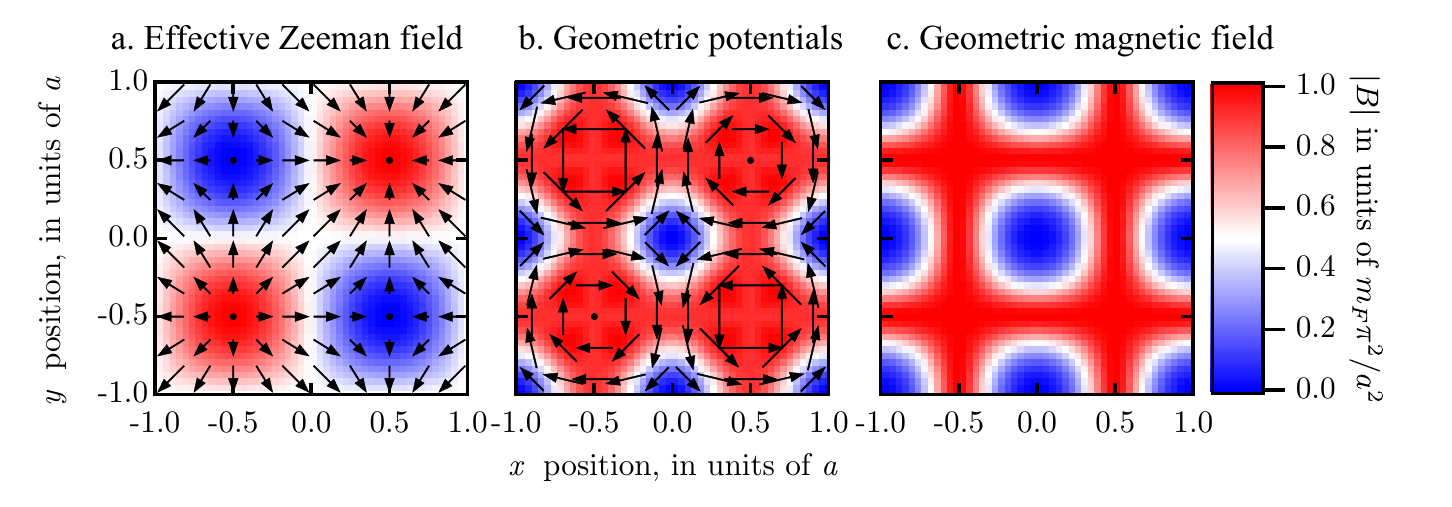}
\end{center}
\caption[Generic Geometry]{Rectangular coordinate flux lattice computed for $\beta=1$.  (a) Position dependent effective Zeeman magnetic field ${\boldsymbol\Omega}$: the vectors denote the components in the $\ex$-$\ey$ plane and the color depicts the $\ez$ component.  (b) Geometric potentials.  The color indicates the geometric scalar potential $W({\bf r})$, and the arrows denote the vector potential $\Ageo({\bf r})$.  (c) Magnitude of effective geometric magnetic field $\Bgeo$ along $\ez$.  Notice that regions of largest $\Bgeo$ correspond to the maxima of $W({\bf r})$. }
\label{fig_ScalarVector}
\end{figure}

Using Eq.~\eqref{eq:B-alter}, one arrives at the explicit result for
the magnetic flux density
\begin{align}
\Bgeo(\mathbf{r})=\hbar m_F\left(\frac{\pi}{a}\right)^2\frac{\beta\left(f_{x}^{2}f_{y}^{2}-1\right)}{\left[f_{x}^{2}+f_{y}^{2}+\beta^2 g_x^2 g_y^2\right] ^{3/2}}\mathbf{e}_{z},\label{eq:B-Square}
\end{align}
where $f_{u}=\cos(u^\prime)$, $g_{u}=\sin(u^\prime)$, and $\beta=\Omega_{\parallel}/\Omega_{\bot}$.  Equation~\eqref{eq:B-Square} explicitly demonstrates that the magnetic flux, while non-uniform,
is non-staggered, and its profile can be tailored by changing the
ratio of the Rabi frequency amplitudes $\beta$. It is evident that
the magnetic flux is zero at the potential minima $x_{n}=na$ and
$y_{m}=ma$ for finite values of $\beta$. For $\beta=1$, Eq.~\eqref{eq:B-Square}
is equivalent to a result obtained independently by J. Dalibard~\cite{Dalibard:unpubl-2011}. 

\begin{figure}[tb!]
\begin{center}
\includegraphics[width=4.5in]{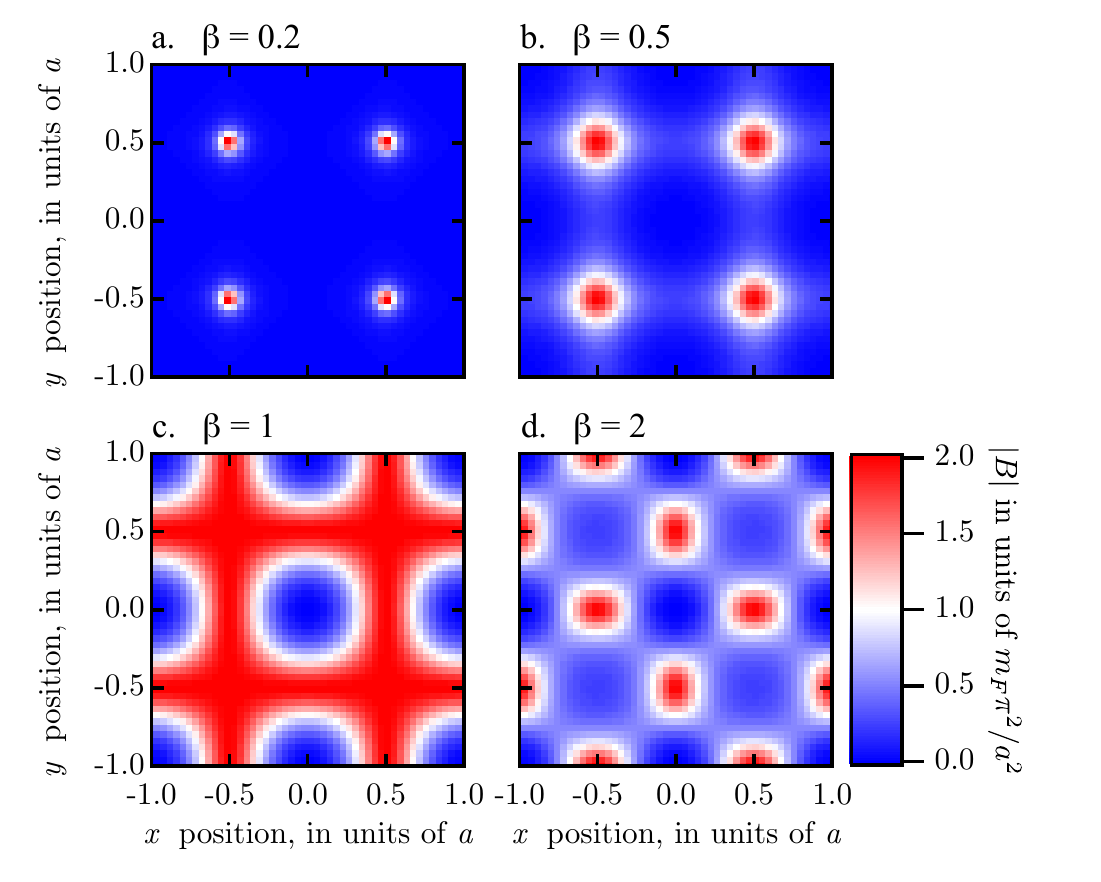}
\end{center}
\caption[Dependence of gauge field on $\beta$]{Dependence of geometric gauge field $\Bgeo({\mathbf r})$ on $\beta$. (a) $\beta=0.2$ (b) $\beta=0.5$ (c) $\beta=1.0$ (d) $\beta=2.0$}
\label{fig_beta}
\end{figure}

Figure~\ref{fig_ScalarVector} displays the spatial distribution of the effective Zeeman field $\boldsymbol{\Omega}$, the geometric vector and scalar potentials $\Ageo$ and $W$, as well as the geometric magnetic field for $\beta=1$. Figure~\ref{fig_beta} presents the geometric magnetic field $\Bgeo$ for various values of $\beta$ showing that the most uniform magnetic field is reached for $\beta=1$.     

\subsection{Scalar potential}\label{sec:Scalar}

Assuming $\Omega_{\parallel}=\Omega_{\bot}$, the second term entering
the scalar potential [Eq.~\eqref{eq:W-result}] is
\begin{equation}
\sin^2\theta\left(\boldsymbol{\nabla}\phi\right)^{2}=\left(\frac{\pi}{a}\right)^{2}\frac{f_{x}^{2}+f_{y}^{2}-2f_{x}^{2}f_{y}^{2}}{\left(1+f_{x}^{2}f_{y}^{2}\right)\left(f_{x}^{2}+f_{y}^{2}\right)}.\label{eq:sin-nabla-phi-square-1}
\end{equation}
It is evident that $\sin^2\theta\left(\boldsymbol{\nabla}\phi\right)^{2}$
is zero at the minima of the adiabatic energy $x_{n}=na$ and $y_{m}=ma$.
Additionally, it equals $\left(\pi/a\right)^{2}$ at the maxima
of the adiabatic energy $x_{n,max}=na+a/2$ and $y_{m,max}=ma+a/2$. 
This part of the scalar potential behaves similar to the adiabatic
energy $E_{-}\left(x,y\right)$, thus increasing the energy maxima
by $\hbar^2\pi^2/8 m a^2$.

The first term entering the scalar potential [Eq.~\eqref{eq:W-result}] is
\begin{equation}
\left(\boldsymbol{\nabla}\theta\right)^{2}=\left(\frac{\pi}{a}\right)^{2}\frac{g_x^2 f_y^2(1+f_x^2)^2 + g_y^2 f_x^2(1+f_y^2)^2}{\left(1+f_{x}^{2}f_{y}^{2}\right)\left(f_{x}^{2}+f_{y}^{2}\right)},\label{eq:grad-theta-square}
\end{equation}
The gradient $\left(\boldsymbol{\nabla}\theta\right)^{2}$ is zero at
the at the minima of the of the adiabatic energy where $g_x = g_y =0$, but is
equal to $\left(\pi/a\right)^{2}$ if $f_x = f_y =0$, i.e. at the center of each plaquette thus raising the potential there.

In this way, the geometric scalar potential is given by Eq.~\eqref{eq:W-result} together with Eqs.~\eqref{eq:sin-nabla-phi-square-1}-\eqref{eq:grad-theta-square}.
It is zero at the corners of a plaquette and reaches its maximum values
at the center of the plaquette, thus behaving similar to the effective magnetic
field,  as is evident in Fig.~\ref{fig_ScalarVector}. The scalar potential thus repels atoms from the area of high magnetic field at the center of the plaquette.

\section{Alkali atoms and light shifts\label{sec:alkali}}

In this Section, we shall first demonstrate a possible way to engineer the state-independent potential  $U(\mathbf{r})$ together with the state-dependent potential $\hat{M}$ featured in the general atomic Hamiltonian given by Eq.~\eqref{eq:hamiltonian1}. Subsequently we shall analyze atom-light configurations providing the square optical flux lattices. 

Let us consider a system of ultracold alkali atoms in their electronic ground state manifold illuminated by one or several laser fields which non-resonantly couple the ground states with the lowest electronic excited states.  In the presence of an external magnetic field (yet without including the contributions due to the laser fields), the Hamiltonian for the atomic ground state manifold is  
\begin{align*}
H_0 =& H_k + A_{\rm hf} \hat{\mathbf I}\!\cdot\!\hat{\mathbf J} + \frac{\mu_B}{\hbar}{\mathbf B}\!\cdot\!\left(g_J\hat{\mathbf J} + g_I\hat{\mathbf I}\right).
\end{align*}
where $H_k={\mathbf p}^2/2m$ is the kinetic contribution to the Hamiltonian; $A_{\rm hf}$ is the magnetic dipole hyperfine coefficient; and $\mu_B$ is the Bohr magneton.  The Zeeman term includes separate contributions from $\hat{\mathbf J}=\hat{\mathbf L}+\hat{\mathbf S}$ (the sum of the orbital $\hat{\mathbf L}$ and electronic spin $\hat{\mathbf S}$ angular momentum) and the nuclear angular momentum $\hat{\mathbf I}$, along with their respective Land\'e-$g$ factors.  We next consider the additional contributions to the atomic Hamiltonian resulting with off-resonant interaction with laser fields. 

As was observed in Refs.~\cite{Deutsch1998,Dudarev2004,Sebby-Strabley2006}, conventional spin independent (scalar, $U_s$) optical potentials acquire additional spin-dependent terms near atomic resonance: the rank-1 (vector, $U_v$) and rank-2 tensor light shifts~\cite{Deutsch1998}.  For the alkali atoms, adiabatic elimination of the excited states labeled by $j =1/2$ (D1) and $j=3/2$ (D2) yields an effective atom-light coupling Hamiltonian for the ground state atoms (with $j=1/2$):
\begin{align*}
H_L&=\left[u_s({\mathbf E}^*\!\cdot\!{\mathbf E}) + \frac{iu_v({\mathbf E}^*\!\times\!{\mathbf E})}{\hbar} \cdot{\mathbf J}\right]
\end{align*}
The rank-2 term is negligible for the parameters of interest and hence is not included in $H_L$. Here $\mathbf E$ is the optical electric field; $u_v=-2u_s\Delta_{\rm FS}/3(\omega-\omega_0)$ determines the vector light shift;  $\Delta_{\rm FS} = \omega_{3/2}-\omega_{1/2}$ is the fine-structure splitting; $\hbar\omega_{1/2}$ and $\hbar\omega_{3/2}$ are the D1 and D2 transition energies; and $\omega_0=(2\omega_{1/2}+\omega_{3/2})/3$ is a suitable average.  $u_s$ sets the scale of the light shift and proportional to the atoms ac polarizability.  
 
The contributions from the scalar and vector light shifts featured in  $H_L$ can be independently specified with informed choices of laser frequency $\omega$ and intensity.  Evidently, the vector light shift is a contribution to the total Hamiltonian acting like an effective magnetic field
\begin{align*}
{\mathbf B}_{\rm eff} &= \frac{iu_v({\mathbf E}^*\!\times\!{\mathbf E})}{\mu_B\g_J}
\end{align*}
that acts on $\hat{\mathbf J}$ and not the nuclear spin $\hat{\mathbf I}$.  Instead of using the full Breit-Rabi equation~\cite{Breit1931} for the Zeeman energies, we assume that the Zeeman shifts are small in comparison with the hyperfine splitting -- the linear, or anomalous, Zeeman regime -- in which case, the effective Hamiltonian for a single manifold of total angular momentum $\hat{\mathbf F}=\hat{\mathbf J} + \hat{\mathbf I}$ states is
\begin{align*}
H_0 + H_L =&\ u_s({\mathbf E}^*\!\cdot\!{\mathbf E}) + \frac{\mu_B g_F}{\hbar}\left({\mathbf B} + {\mathbf B}_{\rm eff}\right)\!\cdot\!\hat{\mathbf F} + \frac{A_{\rm hf}}{2}\left(\hat{\mathbf F}^2 - \hat{\mathbf J}^2 - \hat{\mathbf I}^2\right),
\end{align*}
Notice that ${\mathbf B}_{\rm eff}$ acts as a true magnetic field and adds vectorially with ${\mathbf B}$, and since $|g_I/g_J|\simeq0.0005$ in the alkali atoms, we safely neglected a contribution $-\mu_B g_I {\mathbf B}_{\rm eff}\cdot\hat{\mathbf{I}}/ \hbar$ to the atomic Hamiltonian.   
 We also introduced the hyperfine Land\'e g-factor
\begin{align*}
g_F&= g_J\frac{f(f+1) - i(i+1) + j(j+1)}{2 f (f+1)}.
\end{align*}
In $\Rb87$'s lowest energy manifold with $f=1$, for which $j=1/2$ and $i=3/2$, we get $g_F = -g_J/4\approx-1/2$.  In the following, we always consider a single angular momentum manifold labeled by $f$, and select it's energy at zero field as the zero of energy.

\subsection{Bichromatic light field}

\begin{figure}[t!]
\begin{center}
\includegraphics[width=3.3in]{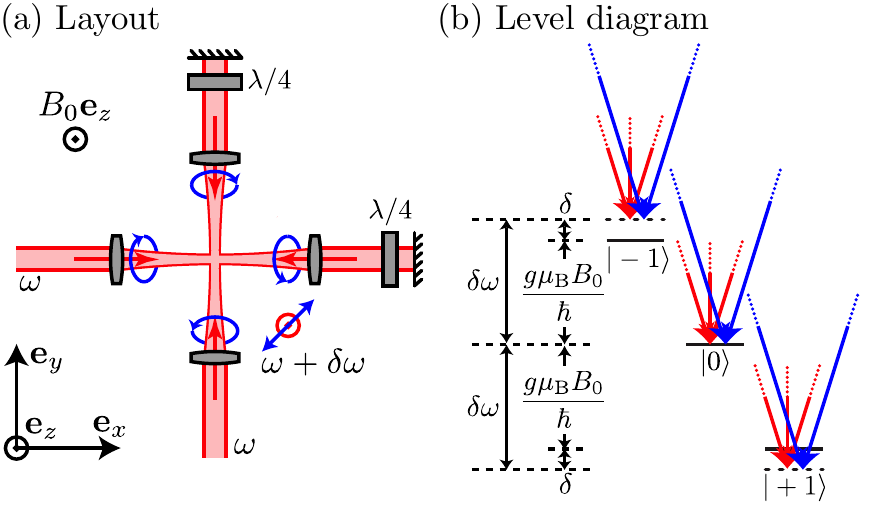}
\end{center}
\caption[Proposed experimental geometry]{Proposed experimental geometry.  (a) Laser geometry for creating flux lattice showing the four circularly polarized beams in the $\ex$-$\ey$ plane with frequency $\omega$ along with the linearly polarized beam traveling along $\ez$ with frequency $\omega+\delta\omega$.  (b) Physical level diagram for three-level total angular momentum $f=1$ case with $m_F$ states labeled, as is applicable for the common alkali atoms$\Li7$, $\Na23$, $^{39}{\rm K}$, $\K41$, and $\Rb87$.  For reference, the diagram shows the decomposition of these optical fields into $\sigma_{\pm}$ and $\pi$, but as discussed in the text, this is not an overly useful way of considering this problem.}
\label{fig_layout}
\end{figure}

By combining state-dependent optical lattices along with ``Raman coupling lattices,'' it is possible to create lattice potentials with large, non-staggered, artificial magnetic fields~\cite{Cooper2011a} even for alkali atoms~\cite{Cooper2011}.   Consider an ensemble of ultracold atoms subjected to a magnetic field ${\mathbf B} = B_0 \ez$. The atoms are illuminated by several lasers with frequencies $\omega$ and $\omega+\delta\omega$, where $\delta\omega\approx |g_F \mu_B B_0/\hbar|$ differs by a small detuning $\delta=g_F \mu_B B_0/\hbar-\delta\omega$ from the linear Zeeman shift between $m_F$ states (where $|\delta|\ll\delta\omega$).  In this case, the complex electric field ${\mathbf E} = {\mathbf E}_{\omega_-}\exp(-i\omega t) + {\mathbf E}_{\omega_+}\exp\left[-i(\omega+\delta\omega) t\right]$ contributes to the combined magnetic field, giving
\begin{align*}
{\mathbf B} + {\mathbf B}_{\rm eff} =& B_0 \ez + \frac{i u_v}{\mu_{\rm B} g_J}\big[\left({\mathbf E}_{\omega_-}^*\!\times\!{\mathbf E}_{\omega_-}\right) + \left({\mathbf E}^*_{\omega_+}\!\times\!{\mathbf E}_{\omega_+}\right) \\
& + \left({\mathbf E}_{\omega_-}^*\!\times\!{\mathbf E}_{\omega_+}\right)e^{-i\delta\omega t} 
+ \left({\mathbf E}^*_{\omega_+}\!\times\!{\mathbf E}_{\omega_-}\right)e^{i\delta\omega t}
\big]\,.
\end{align*} 
The first two terms of $ {\mathbf B}_{\rm eff}$ add to the static bias field $B_0 \ez$, and the remaining two time-dependent terms describe transitions between different $m_F$ levels.  Provided $B_0\gg\left|{\mathbf B}_{\rm eff}\right|$ and $\delta\omega$ are large compared to the kinetic energy scales, the Hamiltonian can be simplified by time-averaging to zero the time-dependent terms in the scalar light shift and making the rotating wave approximation (RWA) to eliminate the time-dependence of the coupling fields.  The resulting contribution to the Hamiltonian
\begin{align}
\hat H_{\rm RWA} &= U({\mathbf r}) \hat 1 + {\boldsymbol \Omega}\!\cdot\!\hat{\mathbf F},\label{RWA_Hamiltonian}
\end{align}
takes the form of Eq.~\eqref{eq:hamiltonian1} once we identify the scalar potential
\begin{align}
U({\mathbf r}) &= u_s \left({\mathbf E}_{\omega_-}^*\!\cdot\!{\mathbf E}_{\omega_-} + {\mathbf E}^*_{\omega_+}\!\cdot\!{\mathbf E}_{\omega_+}\right)
\end{align}
and the RWA effective magnetic field
\begin{align}
{\boldsymbol \Omega} =& \left[\delta + i \frac{u_v}{\hbar}\left({\mathbf E}_{\omega_-}^*\!\times\!{\mathbf E}_{\omega_-}+{\mathbf E}^*_{\omega_+}\!\times\!{\mathbf E}_{\omega_+}\right)\!\cdot\!\ez\right]\ez\nonumber\\
&- \frac{u_v}{\hbar}{\rm Im}\left[\left({\mathbf E}_{\omega_-}^*\!\times\!{\mathbf E}_{\omega_+}\right)\!\cdot\!\left(\ex-i\ey\right)\right]\ex \\
&-\frac{u_v}{\hbar}{\rm Re}\left[\left({\mathbf E}_{\omega_-}^*\!\times\!{\mathbf E}_{\omega_+}\right)\!\cdot\!\left(\ex-i\ey\right)\right]\ey\nonumber.
\end{align}
This expression is valid for $g_F>0$ (for $g_F<0$ the sign of the $\ex$ and $i\ey$ terms would both be  positive, owing to selecting the opposite complex terms in the RWA).
The final form of this effective coupling shows that, while it is related to the initial vector light shifts, ${\mathbf \Omega}$ is composed of both static and resonant couplings in a way that goes beyond the restrictive ${\mathbf B}_{\rm eff} \propto i{\mathbf E}^*\!\times\!{\mathbf E}$ form.  This enables flux lattices in the alkali atoms.

Importantly for practical flux-lattice configurations, $\Omega_z$ depends both on the static magnetic field and also the component of ${\mathbf B}_{\rm eff}$ along ${\bf e}_z$.  For practical considerations it is undesirable that the resonance condition be a function of the laser intensity, so we seek solutions without a contribution from this term.

\subsubsection{Two Raman beams}\label{sec:TwoRaman}

First consider the straightforward example of the two counter propagating Raman beams used in existing experiments~\cite{Lin2009a,Lin2009b,Lin2011,Chen2012,Wang2012,Cheuk2012}.  In this simple case:
\begin{align*}
{\mathbf E}_{\omega_-} &= E e^{i \kr x} \ey,  & {\rm and} &&
{\mathbf E}_{\omega_+} &= E e^{-i \kr x} \ez
\end{align*}
describing the electric field of two lasers counterpropagating along $\ex$ with equal intensities and crossed linear polarization, where $\kr=2\pi/\lambda$ is the single photon recoil wave vector, and $\Er=\hbar^2 \kr^2/2m$ is the associated recoil energy.
The resulting scalar light shift $U({\mathbf r})$ and the effective magnetic field ${\boldsymbol \Omega}$ describing the vector light shift are
\begin{align*}
U({\mathbf r}) &= u_s \left({\mathbf E}_{\omega_-}^*\!\cdot\!{\mathbf E}_{\omega_-} + {\mathbf E}^*_{\omega_+}\!\cdot\!{\mathbf E}_{\omega_+}\right) =2 u_s E^2 \\
{\boldsymbol \Omega} &= \delta \ez + \Omega_R\left[\sin\left(2 \kr x\right)\ex - \cos\left(2\kr x\right)\ey\right],
\end{align*}
where $\Omega_R = u_v E^2/\hbar$.  These describe a constant scalar light shift along with a spatially rotating effective magnetic field, as discussed in Ref.~\cite{Lin2011} which produced an artificial spin-orbit coupling, and in different notation, is equivalent to the proposal of Ref.~\cite{Juzeliunas2006}.  Because this Hamiltonian is only invariant under spatial translations with primitive vector ${\bf u} = \pi/\kr\ex$, it would be expected to describe a periodic lattice.  However, transforming the complete Hamiltonian according to the rotation $\hat U(x) \hat H \hat U^\dagger(x)$, with $\hat U(x) = \exp\left[i\hat F_z\left(2\kr x - \pi/2\right)/\hbar\right]$, completely removes the spatial periodicity.  Instead, the transformed Hamiltonian becomes
\begin{align*}
\hat H &= \frac{\hbar^2}{2m}\left(\hat k - 2\kr \hat F_z/\hbar\right)^2 + U({\mathbf r})  \hat 1 + \delta \hat F_z + \Omega_R \hat F_x,
\end{align*}
in which the position dependence has vanished from coupling vector $\delta \ez + \Omega_R \ey$.  In this example, all spatial dependance (including the initial lattice structure) has been eliminated from the Hamiltonian in exchange for a matrix valued (though Abelian) gauge field.  An additional spatially uniform radio-frequency magnetic field added to the mix forces the spatial structure to remain, creating a composite lattice potential~\cite{Jimenez-Garcia2012}.

\subsubsection{Flux lattice configuration}\label{sec:Generation-fo-the}

Next, we analyze the configuration shown in Fig.~\ref{fig_layout} where four beams with angular frequency $\omega$ intersecting at right angles in the $\ex$-$\ey$ plane, are joined by a fifth beam with angular frequency $\omega+\delta\omega$ traveling along $\ez$.  The total electric field from these five beams are
\begin{align*}
{\mathbf E}_{\omega_-} =& {\mathbf E}_{x^+} +  {\mathbf E}_{x^-} +  {\mathbf E}_{y^+} +  {\mathbf E}_{y^-}, &{\rm and} && {\mathbf E}_{\omega_+} &= {\mathbf E}_{z}
\end{align*}
where
\begin{align*}
{\mathbf E}_{x^+} =& E_{xy} \left(e^{-i\phi/2}\cos\theta_p\ez+e^{i\phi/2}\sin\theta_p\ey\right) e^{i\delta\phi_x/2}e^{i\delta\phi_{xy}/2} e^{i \kr x}\\
{\mathbf E}_{x^-} =& E_{xy} \left(e^{-i\phi/2}\cos\theta_p\ez-e^{i\phi/2}\sin\theta_p\ey\right) e^{-i\delta\phi_x/2} e^{i\delta\phi_{xy}/2}e^{-i \kr x}\\
{\mathbf E}_{y^+} =& E_{xy} \left(e^{i\phi/2}\cos\theta_p\ez-e^{-i\phi/2}\sin\theta_p\ex\right) e^{i\delta\phi_y/2} e^{-i\delta\phi_{xy}/2}e^{i \kr y}\\
{\mathbf E}_{y^-} =& E_{xy} \left(e^{i\phi/2}\cos\theta_p\ez+e^{-i\phi/2}\sin\theta_p\ex\right) e^{-i\delta\phi_y/2} e^{-i\delta\phi_{xy}/2}e^{-i \kr y}\\
{\mathbf E}_{z} =& \frac{E_z}{\sqrt{2}}  \left(\ex+\ey\right) e^{i \kr z}.
\end{align*}
In this complicated set of fields, $\phi$ describes the ellipticity of the lasers traveling in the $\ex$-$\ey$ plane, the major axes of which are tipped by an angle $\theta_p$ from vertical.  When $\phi=\pi/2$ all four beams are right-hand circular polarized.   $\delta\phi_{x}$ and $\delta\phi_{y}$ describe relative phase differences between the forward- and counter-propagating beams along $\ex$ and $\ey$ respectively; and lastly, $\delta\phi_{xy}$ is an overall phase difference between the beams traveling along $\ex$ and those traveling along $\ey$ (a similar phase difference $\delta\phi_{z}$ exists between the $\ex$-$\ey$ beams and the $\ez$ beam, however, it amounts to simply displacing the system along $\ez$). 

For this set of fields, the scalar light shift [neglecting a $u_s(4 E_{xy}^2 + E_z^2)$ energy offset] of is
\begin{align}
U({\bf r}) &= U_\perp \left[\cos(2x^\prime)+\cos(2y^\prime)\right] + U_\parallel \cos x^\prime \cos y^\prime,\label{eq:PhysicalScalar}
\end{align}
where we have introduced the scalar energies $U_\perp = 2 u_s E_{xy}^2 \cos\left(2\theta_p\right)$ and $U_\parallel=8 u_s E_{xy}^2 \cos^2\theta_p\cos(2\varphi_-)$, with $\varphi_\pm = (\delta\phi_{xy}\pm\phi)/2$.  The RWA effective magnetic field becomes
\begin{align}
{\boldsymbol \Omega} &= \Omega_\perp\left[
	\cos\left(x^\prime\right)\sin\left(z^\prime-\varphi_-\right) + 
	\cos\left(y^\prime\right)\sin\left(z^\prime +\varphi_-\right)
\right]\ex\nonumber\\
 &+ \Omega_\perp\left[
	\cos\left(x^\prime\right)\cos\left(z^\prime-\varphi_-\right) + 
	\cos\left(y^\prime\right)\cos\left(z^\prime+\varphi_-\right)
\right]\ey\nonumber\\
 &+ \Omega_\parallel\left[\sin(x^\prime)\sin(y^\prime) + \tilde\delta\right] \ez.\label{eq:PhysicalZeeman}
\end{align}
We  defined $\Omega_\perp = 2 u_v E_{xy}E_z\cos\theta_p/\hbar$ and $\Omega_\parallel = 4 u_v E_{xy}^2\sin(2\varphi_+)\sin^2\theta_p/\hbar$ and introduced a dimensionless detuning $\tilde\delta=\delta/\hbar\Omega_\parallel$. (Here, ${\mathbf E}_{z}$ is linearly polarized, so it does not have any contributions to $\delta$, as would be the case for a circularly polarized beam~\cite{Cooper2011a}.) In these expressions, we made the simplifying replacements $x^\prime = \kr x - \delta\phi_x/2$, $y^\prime = \kr y - \delta\phi_y/2$, and $z^\prime = \kr z + \pi/4$.  These show that the phase-differences between beams traveling along $\ex$ and $\ey$ give rise only to effective spatial displacements leaving the topology of the lattice unchanged; in contrast, the phase difference between the $\ex$ and $\ey$ lasers $\delta\phi_{xy}$ fundamentally change the coupling. 

Somewhat more subtly, transforming the complete Hamiltonian according to the unitary rotation $\hat U(x) \hat H \hat U^\dagger(x)$, with $\hat U(x) = \exp\left(i\hat F_z z^\prime/\hbar\right)$ completely eliminates the $z$-dependence from the Hamiltonian, but as in Sec.~\ref{sec:TwoRaman}, introduces a gauge field $k_L \hat F_z$ for motion along $\ez$.   Under this transformation the effective Zeeman term becomes
\begin{align*}
\hat U(x) \left[\hat {\mathbf F}\cdot\boldsymbol\Omega(z^\prime)\right] \hat U^\dagger(x) &= \hat {\mathbf F}\cdot\boldsymbol\Omega(z^\prime=0).
\end{align*}
Therefore the Hamiltonian separates into a sum of independent contributions for motion along $\ez$ and motion in the $\ex$-$\ey$ plane; without loss of generality, we take $z^\prime = \pi/4$.  The expression for ${\boldsymbol \Omega}$ then reduces to that of Eq.~\eqref{eq:Omega-xyz-Square} for the physical parameters $\phi = \pi/2$ (circularly polarized beams in the $\ex$-$\ey$ plane), $\phi_{xy}=0$, and $\tilde\delta=0$.

With the replacement $\beta = \Omega_\parallel /\Omega_\perp$, the resulting adiabatic orbital field is
\begin{align}
\Bgeo({\mathbf r}) &= \hbar m_f \left(\frac{\pi}{a}\right)^2\frac{\beta\left(f_x^2 f_y^2-1-\tilde\delta g_x g_y\right)\sin\left(-2\varphi_-\right)}{\left[f_x^2+f_y^2+2f_x f_y\cos(2\varphi_-)+\beta^2\left(\tilde\delta + g_x g_y\right)^2\right]^{3/2}}\ez. \label{eq:B-Square-Raman}
\end{align}
This implies that practical implementations of flux lattices require active stabilization of the phase between beams traveling along $\ex$ and $\ey$, but not the $\ez$ beam.  For the choice $\varphi_- = -\pi/4$ and $\tilde\delta=0$ this reduces to Eq.~\eqref{eq:B-Square}.

Given the dependance of $\Bgeo({\mathbf r})$ on so many parameters, we now consider the first order sensitivity to perturbations in $\tilde\delta=\Delta\tilde\delta$ and $\varphi_- =-\pi/4 + \Delta\varphi_-$; since changes in phase sum $\varphi_+ = \pi/4 + \Delta\varphi_0$ enter into $\Omega_\parallel$ quadratically, they may be neglected at first order.  Additionally, the polarization angle $\theta_P$ is generally static in the lab, and an imperfect setting can be accounted for by changing the intensity of the Raman beams.  We find the scalar light shift is unchanged, but the effective coupling becomes
\begin{align}
{\mathbf \Omega} &=
\Omega_\perp \left[\cos\left(x^\prime\right) + \Delta\varphi_-\cos\left(y^\prime\right)\right]\ex\nonumber\\
 &+ \Omega_\perp \left[-\Delta\varphi_-\cos\left(x^\prime\right) + \cos\left(y^\prime\right)\right]\ey\\
 &+ \Omega_\parallel \left[ \sin(x^\prime)\sin(y^\prime) + \Delta\tilde\delta\right] \ez.\nonumber
\end{align}
Given this, it is surprising but delightful, that we arrive at an orbital field which is un-altered at first order.  We observe that $\Delta\tilde\delta$ usually results from noise in the magnetic field; here this noise must be small compared to the coupling strength $\Omega_\parallel$, not the generally much smaller width of the Bloch bands.

\subsection{Band structure and Chern numbers}

The adiabatic arguments show that flux lattices give rise to large orbital magnetic fields with non-zero average.  As we learned above, the spatial locations with largest magnetic field are also associated with a repulsive adiabatic scalar potential $W({\mathbf r})$, suggesting that without a compensating term from the scalar potential $U({\mathbf r})$ the magnetic field might not be important for atoms in the lowest bands.  To address this question, we studied the resulting band structure and identified when the Bloch bands have non-zero Chern number, in analogy with the band structure of charged particles in a magnetic field.  We directly compute the band structure from potential terms in Eq.~\eqref{RWA_Hamiltonian} combined with the contribution from the usual kinetic energy term, and then extract the Chern numbers using the prescription in Ref.~\cite{Fukui2005}.

\begin{figure}[t]
\begin{center}
\includegraphics[width=4.25in]{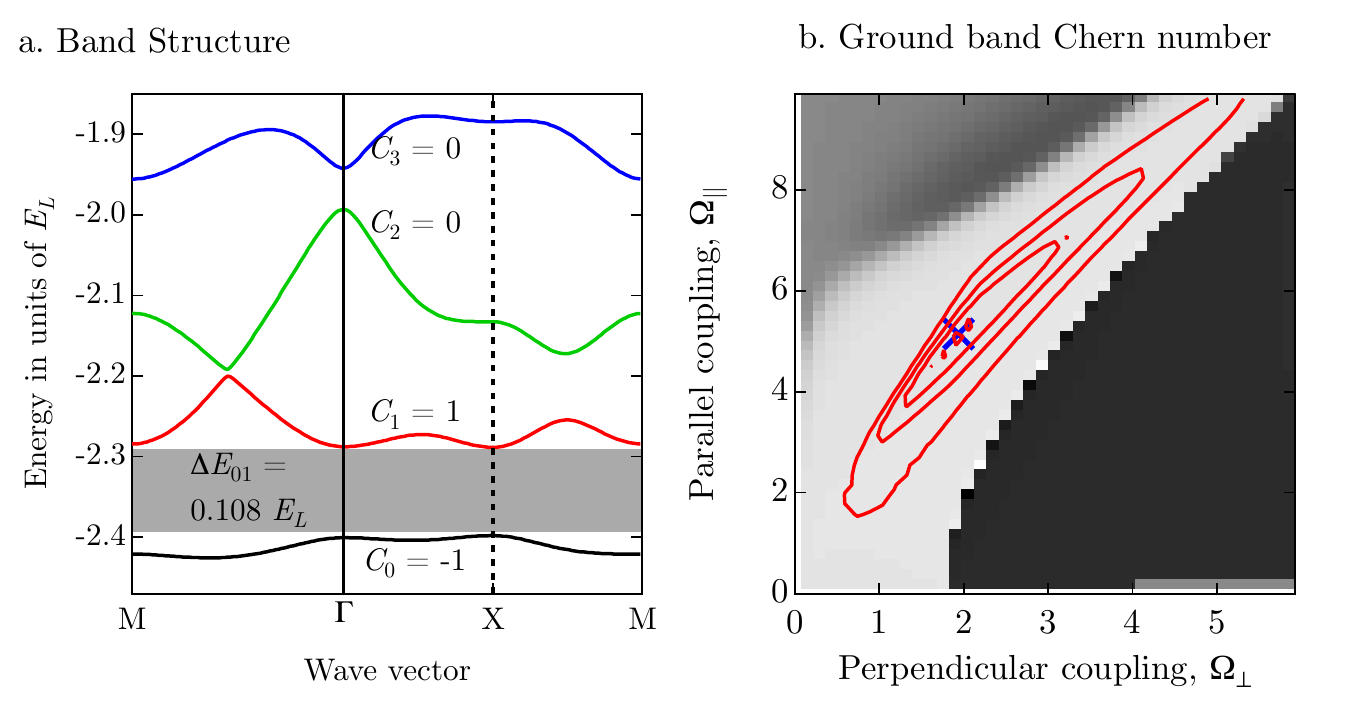}
\end{center}
\caption[Results]{Band structure.  (a) Band structure for the lowest four bands showing the Chern numbers $C_n$, and the modest energy gap $\Delta E_{01}$ between the ground and first excited band.  These were computed for $\Omega_\perp = 1.905\El$, $\Omega_\parallel = 5.1\El$, $U_\perp = -1.95\El$, $U_\parallel = 0$, $\delta = 0$ and the phase $\phi_- = -\pi/4$.  (b) Ground band Chern number as a function of $\Omega_\perp$ and $\Omega_\parallel$ for the same $U_\perp$ and $U_\parallel$ as above.  The red contours mark $\Delta E_{01}$ in this plane, showing the modest parameter region where it is non-negligible, and the blue cross locates the maximum gap where the band structure of (a) was computed.}
\label{fig_results}
\end{figure}

The Hamiltonian described by  Eq.~\eqref{RWA_Hamiltonian} has apparent primitive lattice vectors ${\mathbf u}_1 = 2\pi/\kr\ex$ and ${\mathbf u}_2 = 2\pi/\kr\ey$ (each of these is twice larger than usual for a lattice formed by retro-reflected lasers).  To compute the band structure in the most simple manner we first rotated the coordinate system in Eqs.~\eqref{eq:PhysicalScalar} and ~\eqref{eq:PhysicalZeeman} by $\pi/4$ in the $\ex$-$\ey$ plane and defined scaled coordinates $x^\pprime = y^\prime+x^\prime$ and $y^\pprime=y^\prime-x^\prime$.  In analogy with the procedure described in Sec.~\ref{sec:TwoRaman}, we applied a spatially dependent rotation
\begin{align*}
U({\mathbf r}) &= \exp\left[\frac{i\left(x^\pprime+y^\pprime\right)\hat F_z}{2\hbar}\right] = \exp\left[\frac{i\kr y \hat F_z}{\hbar}\right]
\end{align*}
which introduced a gauge term in the kinetic energy.  In the example given in Sec.~\ref{sec:TwoRaman}, this process completely removed the Hamiltonian's spatial dependence; here it does not, but the area of the unit cell is halved [the primitive lattice vectors ${\mathbf u}_1 = \pi/2\kr(\ex+\ey)$ and ${\mathbf u}_2 = \pi/2\kr(-\ex+\ey)$ expressed in the initial coordinate system are reduced in magnitude by a factor of $1/\sqrt{2}$], doubling the area of the Brillouin zone~\cite{Cooper2011a}.  The resulting Hamiltonian has contributions
\begin{align}
H_k =& \frac{\hbar^2}{2m}\left[\left(k_x - \kl \hat F_z/2\right)^2+\left(k_y - \kl \hat F_z/2\right)^2\right]\nonumber\\
U({\mathbf r}) =& U_\perp\left[\cos(x^\pprime+y^\pprime)+\cos(x^\pprime-y^\pprime)\right] + \frac{U_\parallel}{2} \left[\cos\left(x^\pprime\right) + \cos\left(y^\pprime\right)\right]\\
\hat{\mathbf F}\cdot\boldsymbol{\Omega} =& \hat F_z \Omega_z + \hat F_+ \Omega_- + \hat F_- \Omega_+
\nonumber.
\end{align}
Where $\hat F_\pm = \hat F_x \pm i \hat F_y$ are the usual angular momentum raising and lower operators; the coupling expressed in the helicity basis is quite simple with
\begin{align*}
\Omega_z &= \frac{\Omega_\parallel}{2}\left[-\cos(x^\pprime) + \cos(y^\pprime) + \tilde\delta\right] \\
\Omega_+ = \Omega_-^\dagger &= \frac{i\Omega_\perp}{4} 
\left[e^{i\left(x^\pprime+\varphi_-\right)} + e^{i\left(y^\pprime+\varphi_-\right)} + e^{i\left(x^\pprime+y^\pprime-\varphi_-\right)}+e^{-i\varphi_-}\right].
\end{align*}
All of these expressions fully respect translational symmetry in the reduced unit cell whose reciprocal lattice vectors have magnitude $\kl = \sqrt{2}\kr$, and a recoil energy $\El = 2\Er$.  From this, computation of the band structure and it's Chern numbers is straightforward.

Figure~\ref{fig_results} depicts the outcome of this computation for an optimally chosen parameter set (values given in the caption).  While the Chern number of the lowest band $C_0$ is non-zero over a wide range of parameters, but red contours illustrate the most significant limitation for practical implementation of these flux lattices:  the relatively small gap between the ground and first excited band $\Delta E_{01}$.  For our optimal parameter set, we find a maximal gap of just $\Delta E_{01}=0.107\El=0.214\Er$, far less than the $U\approx 1\Er$ on-site interaction energy in typical 3D optical lattices (a slight improvement is possibly by tuning the quadratic Zeeman term which was absent in these computations).  This implies that interactions will hybridize several of the lowest bands in a way that cannot be described as a perturbation of the lowest band as is possible in conventional optical lattices where $\Delta E_{01}\gtrsim10\El$.

\section{Concluding remarks}
We have explored the optical flux lattices produced for ultra-cold atoms
in the radiation field when both the atom-light coupling and the detuning
exhibit an oscillatory behavior. We have analyzed not only the magnetic
flux but also the geometric vector potential generating the flux,
as well as the accompanying geometric scalar potential. We showed how
to deal with the gauge-dependent singularities of the AB type appearing in the vector potentials for the optical flux lattices. We have present a way to calculate the continuous magnetic flux through the elementary cell via the singularities of the vector potential inside the cell. The analysis is illustrated with a square
optical flux lattice. We have presented a way of creating such a lattice
using the Raman transitions induced by a set of properly chosen polarization-dependent
standing waves propagating at a right angle and containing a time-phase
difference.

\section{Acknowledgments}

The authors acknowledge helpful discussions with Jean Dalibard, Nigel Cooper, Tilman Esslinger, Janne Ruostekoski, Julius Ruseckas, Algirdas Mekys and Simonas Grubinskas.  In addition, we appreciate a careful reading by L. J. LeBlanc.  G.J acknowledges the support from the Research Council of Lithuania (Grant No. MIP-082/2012).  I.B.S. acknowledges the financial support of the NSF through the PFC at JQI, and the ARO with funds from both the Atomtronics MURI and the DARPA OLE Program.

\bibliographystyle{unsrt}
\bibliography{FluxLattice}

\end{document}